\makeatletter
\@ifundefined{@parse@version@dash}{%
\def\@parse@version#1{\@parse@version@0#1}
\def\@parse@version@#1/#2/#3#4#5\@nil{%
\@parse@version@dash#1-#2-#3#4\@nil}
\def\@parse@version@dash#1-#2-#3#4#5\@nil{%
  \if\relax#2\relax\else#1\fi#2#3#4 }
}{}
\makeatother

\documentclass[aps,prd,notitlepage,nofootinbib,nobibnotes,superscriptaddress,amsmath,amssymb,preprintnumbers]{revtex4-2}

\usepackage{graphicx} 
\usepackage{dcolumn}
\usepackage{amssymb}
\usepackage{amsmath}
\usepackage[hidelinks]{hyperref}
\usepackage{color}
\usepackage[normalem]{ulem}
\usepackage[utf8]{inputenc}
\usepackage{slashed}

\newcommand{\be}{\begin{equation}}
\newcommand{\ee}{\end{equation}}

\newcommand{\rmd}{\mathrm{d}}
\newcommand{\rme}{\mathrm{e}}
\newcommand{\rmi}{\mathrm{i}}

\newcommand{\calP}{\mathcal{P}}
\newcommand{\calM}{\mathcal{M}}

\newcommand{\calS}{\mathcal{S}}
\newcommand{\calO}{\mathcal{O}}
\newcommand{\calG}{\mathcal{G}}

\newcommand{\calT}{\mathcal{T}}
\newcommand{\calH}{\mathcal{H}}

\newcommand{\rp}{\boldsymbol{r}_{\perp}}

\newcommand{\donp}{\boldsymbol{\delta}_{1\perp}}
\newcommand{\dtwp}{\boldsymbol{\delta}_{2\perp}}
\newcommand{\bp}{\boldsymbol{b}_{\perp}}
\newcommand{\xp}{\boldsymbol{x}_{\perp}}

\newcommand{\vp}{\boldsymbol{v}_{\perp}}
\newcommand{\vonp}{\boldsymbol{v}_{1\perp}}
\newcommand{\vtwp}{\boldsymbol{v}_{2\perp}}
\newcommand{\hvonp}{\hat{\boldsymbol{v}}_{1\perp}}
\newcommand{\hvtwp}{\hat{\boldsymbol{v}}_{2\perp}}

\newcommand{\yp}{\boldsymbol{y}_{\perp}}

\newcommand{\qp}{\boldsymbol{q}_{\perp}}
\newcommand{\qop}{\boldsymbol{q}_{1\perp}}
\newcommand{\Sp}{\boldsymbol{S}_{\perp}}
\newcommand{\pp}{\boldsymbol{p}_{\perp}}

\newcommand{\kp}{\boldsymbol{k}_{\perp}}
\newcommand{\kapp}{\boldsymbol{\kappa}_{\perp}}
\newcommand{\delp}{\boldsymbol{\Delta}_{\perp}}

\newcommand{\khp}{\boldsymbol{k}_{1\perp}}
\newcommand{\kAp}{\boldsymbol{k}_{2\perp}}
\newcommand{\kgp}{\boldsymbol{k}_{g\perp}}

\newcommand{\pd}{\partial}

\begin{document}
\date{\today}
\preprint{ZTF-EP-22-05}

\title{On the odderon mechanism for transverse single spin asymmetry in the Wandzura-Wilczek approximation}
\author{Sanjin Beni\' c}
\affiliation{Department of Physics, Faculty of Science, University of Zagreb, Bijenička c. 32, 10000 Zagreb, Croatia}
\author{Davor Horvati\' c}
\affiliation{Department of Physics, Faculty of Science, University of Zagreb, Bijenička c. 32, 10000 Zagreb, Croatia}
\author{Abhiram Kaushik}
\affiliation{Department of Physics, Faculty of Science, University of Zagreb, Bijenička c. 32, 10000 Zagreb, Croatia}
\author{Eric Andreas Vivoda}
\affiliation{Department of Physics, Faculty of Science, University of Zagreb, Bijenička c. 32, 10000 Zagreb, Croatia}
\begin{abstract} 
We compute the transverse single spin asymmetry in forward $p^\uparrow p \to hX$ and $p^\uparrow A \to hX$ collisions from the odderon mechanism originally suggested by Kovchegov and Sievert \cite{Kovchegov:2012ga}. Working in the hybrid approach of the Color Glass Condensate effective theory we firstly identify the relevant collinear parton distribution function (PDF) of the transversely polarized proton $p^\uparrow$ as the intrinsic twist-3 $g_T(x)$ distribution.
We further argue that the complete polarized cross section also contains contributions from the kinematical and the dynamical twist-3 PDFs, in addition to the intrinsic twist-3 PDF. By restricting to the Wandzura-Wilczek approximation, where the dynamical twist-3 PDFs are dropped, we find that the odderon contribution to the polarized cross section for inclusive hadron production is exactly zero at the next-to-leading order in the strong coupling.
\end{abstract}

%\pacs{...}

\maketitle

\section{Introduction and motivation}

Transverse single spin asymmetry (SSA) \cite{Barone:2001sp,DAlesio:2007bjf,Pitonyak:2016hqh,GrossePerdekamp:2015xdx} is a phenomena associated with azimuthally asymmetric particle production in collisions involving a transversely polarized proton $p^\uparrow$. SSA is characterized by a sine modulation $\boldsymbol{P}_{h\perp} \times \Sp = P_{h\perp} S_\perp \sin(\phi_h  -\phi_S)$. Here $\boldsymbol{P}_{h\perp}$ is the transverse momentum of the produced hadron and $\Sp$ is the spin of the transversely polarized proton. Decades of dedicated measurements have demonstrated its persistence even at the highest collision energies, with the SSA being largest in the forward region of the produced particle, typically a hadron. This is so across different collision systems such as $ep^\uparrow$ and $p^\uparrow p$ but also most recently for $p^\uparrow A$ collisions \cite{PHENIX:2019ouo,STAR:2020grs}.

On the theory front, it is known that a presence of the phase in the cross section is crucial to generate SSA. In the forward region, where the momentum fraction $x$ in the target is small, one naturally expects the phenomena of gluon saturation \cite{Iancu:2003xm,Gelis:2010nm,Kovchegov:2012mbw,Blaizot:2016qgz} to play an important role in determining SSA \cite{Boer:2006rj,Kang:2011ni,Kovchegov:2012ga,Schafer:2014zea,Zhou:2015ima,Hatta:2016wjz,Hatta:2016khv,Benic:2018amn,Kovchegov:2020kxg}. In this work we are revisiting the computation by Kovchegov and Sievert \cite{Kovchegov:2012ga}, where they used the Color Glass Condensate (CGC) effective theory \cite{Iancu:2003xm,Gelis:2010nm,Kovchegov:2012mbw,Blaizot:2016qgz} for gluon saturation, to suggest a new mechanism for SSA. The special property of this mechanism is in supplying the phase by the odderon distribution \cite{Kovchegov:2003dm,Hatta:2005as,Jeon:2005cf}
\be
\calO(\xp,\yp) \equiv \frac{1}{2\rmi N_c} {\rm tr}\left\langle V(\xp)V^\dag(\yp) - V(\yp)V^\dag(\xp) \right\rangle\,,
\label{eq:odderon}
\ee
that is, the imaginary part of the dipole distribution ${\rm tr}\left\langle V(\xp)V^\dag(\yp)\right\rangle/N_c$. Here $V(\xp)$ is a fundamental Wilson line with $\langle \dots\rangle$ denoting the color average. This ``odderon mechanism", as we will refer to it in this work, leads to a substantial $A$ suppression of SSA, $\sim A^{-7/6}$ parametrically \cite{Kovchegov:2012ga}.

In the following Sec.~\ref{sec:gen} we take as a starting point the polarized cross section in the hybrid approach \cite{Schafer:2014zea,Zhou:2015ima,Hatta:2016wjz,Hatta:2016khv} with a transversely polarized proton described by the collinear twist-3 PDFs and the dense target (a nuclei or a proton in the forward collision) by Wilson line correlators arising from the CGC. In the context of the twist-3 parton distribution functions (PDFs) the computation of \cite{Kovchegov:2012ga} is clarified in terms of the $g_T(x)$ distribution. We carefully emphasize, however, that the complete twist-3 hadronic cross section contains additional terms associated with the kinematic twist-3 function $g^{(1)}_{1T}(x)$, that is, the first moment of the worm-gear TMD \cite{Bacchetta:2008xw}, as well as the dynamical Efremov-Teryaev-Qiu-Sterman (ETQS) functions \cite{Efremov:1981sh,Qiu:1998ia}, see e.~g. Eq.~\eqref{eq:wtw3} below. Working in the Wandzura-Wilczek (WW) approximation \cite{Wandzura:1977qf}, that neglects the dynamical twist-3 part of the cross section, our computation firstly confirms that the cross section is proportional to the odderon distribution \eqref{eq:odderon}. However, as we explicitly show in Secs.~\ref{sec:qg}, \ref{sec:qqbar} and \ref{sec:ggg} that, due to the specific form of the resulting hard factor, the odderon contribution to SSA for inclusive hadron production turns out to be exactly zero at the next-to-leading order (NLO) in the strong coupling $\alpha_S$ for all possible partonic channels. In the concluding Sec.~\ref{sec:conc} we also briefly outline several new ways the odderon could appear in SSA after all.

\section{General remarks}
\label{sec:gen}

In the hybrid approach a dilute projectile proton is described using collinear PDFs, while the distributions of the dense target (nuclei, or a proton in forward collisions) are given in terms of Wilson line correlators. To set up our notations we first write down the unpolarized $p A \to h X$ cross section in terms of the familiar twist-2 PDFs and fragmentation functions (FFs). For convenience, this is given in the following way
\be
E_h\frac{\rmd \sigma}{\rmd^3 P_h} \equiv \frac{1}{2 (2\pi)^3} \int \frac{\rmd z_h}{z_h^2} D(z_h)\int \rmd x_p \left\{\frac{1}{2}f(x_p) {\rm Tr}\left[\slashed{P}_p S^{(0)}(p_1)\right]+\frac{1}{2}G(x_p)(-g_{\perp}^{\alpha\beta})S^{(0)}_{\alpha\beta}(p_1)\right\}\,,
\label{eq:xsecun}
\ee
where we have separated out the the twist-2 hadron FF $D(z_h)$ and the twist-2 PDF in the quark (gluon) $f(x_p)$ ($G(x_p)$) initiated channel. The proton and the nucleus move along the light-cone with momenta in the center-of-mass frame given as $P_p^+ = P_A^- = \sqrt{s/2}$, where $s$ is the collision energy squared per nucleon\footnote{Here $P_A^-$ is the center-of-mass momentum per nucleon.}, while $P_h$ is the momenta of the produced hadron $h$. Here and in the following we use the light-cone variables $p^{\pm} = (p^0 \pm p^3)/\sqrt{2}$. Furthermore, $p_1$ is the momentum of the parton moving collinearly with the proton $p_1 = x_p P_p$, and $g_{\perp}^{\alpha\beta} = g^{\alpha\beta} - n^\alpha\bar{n}^\beta - \bar{n}^{\alpha}n^\beta$ with $n^\alpha = \delta^{\alpha -}$, $\bar{n}^\alpha = \delta^{\alpha +}$, is a transverse projector to the physical gluon polarizations. $S^{(0)}(p_1)$ ($S^{(0)}_{\alpha\beta}(p_1)$) is an all-order two-parton scattering kernel in the quark (gluon) initiated channel containing the hard factor and also the target distribution that we will be computing within the CGC hybrid approach. We have absorbed the CGC flux factor $1/2P_p^+$ into the definition of $S^{(0)}(p_1)$.

\subsection{Polarized cross section}

In order to generalize to collisions with a transversely polarized proton, the polarized cross section, $\rmd \Delta \sigma$, is computed up to twist-3 in the polarized proton PDF. We will be restricting here to the usual twist-2 FF, $D(z_h)$ -- the particular contribution arising from twist-3 FFs \cite{Ji:1993vw,Yuan:2009dw} has been computed in $p^\uparrow A$ \cite{Hatta:2016khv}. We are also not considering various pole contributions to $\rmd \Delta\sigma$, see \cite{Hatta:2016wjz}. Our starting point is a separate (non-pole) contribution that has already been discussed in SIDIS \cite{Ratcliffe:1985mp,Benic:2019zvg}. Adapting to the $p^\uparrow A$ computation we have the following gauge invariant all-order expression\footnote{We are using the convention $\epsilon_{0123} = +1 = - \epsilon^{0123}$ and $\gamma_5 = \rmi\gamma^0 \gamma^1 \gamma^2 \gamma^3$.}
\be
\begin{split}
E_h\frac{\rmd \Delta\sigma}{\rmd^3 P_h} &= \frac{1}{2(2\pi)^3}\int\frac{\rmd z_h}{z_h^2} D(z_h)\Bigg\{\frac{M_N}{2}\int \rmd x_p g_T(x_p){\rm Tr}\left[\gamma_5 \slashed{S}_\perp  S^{(0)}(p_1)\right]\\
& + \frac{M_N}{2}\int \rmd x_p g_{1T}^{(1)}(x_p) {\rm Tr}\left[\gamma_5 \slashed{P}_p S_\perp^\lambda \left(\frac{\pd S^{(0)}(k_1)}{\pd k_{1\perp}^\lambda}\right)_{k_1 = p_1}\right]\\
& + \frac{\rmi M_N}{4}\int \rmd x_p \rmd x_p' {\rm Tr}\left[\left(\slashed{P}_p \epsilon^{\bar{n} n \lambda S_\perp}\frac{G_F(x_p,x_p')}{x_p - x_p'} + \rmi \gamma_5 \slashed{P}_p S_\perp^\lambda \frac{\tilde{G}_F(x_p,x_p')}{x_p - x_p'}\right)S^{(1)}_\lambda(x_p P_p, x_p'P_p)\right]\Bigg\}\,.
\end{split}
\label{eq:wtw3}
\ee
The first part of \eqref{eq:wtw3} is arising from the $g_T(x_p)$ distribution function. This is also referred to as a {\it intrinsic} (i. e. $\sim \langle P_p S_\perp | \bar{\psi}\psi |P_p S_\perp\rangle$) contribution. The second part is a {\it kinematical} ($\langle P_p S_\perp | \bar{\psi}\partial\psi | P_p S_\perp \rangle $) contribution that is proportional to $g_{1T}^{(1)}(x_p)$, namely the first moment of the worm-gear TMD\footnote{Instead of $g_{1T}^{(1)}(x)$ sometimes a function $\tilde{g}(x)$ is used \cite{Eguchi:2006qz}, with the relation $\tilde{g}(x) = - 2 g_{1T}^{(1)}(x)$ \cite{Kanazawa:2015ajw}.}. In the third ({\it dynamical} $\sim \langle P_p S_\perp | \bar{\psi}F\psi | P_p S_\perp \rangle $) part we have the ETQS distributions $G_F(x_p,x_p')$ and $\tilde{G}_F(x_p,x_p')$. Note the appearance of the same two-parton scattering kernel $S^{(0)}(p_1)$ as in \eqref{eq:xsecun}. To compute the cross section we also need its finite-$k_{1\perp}$ variant $S^{(0)}(k_1)$, as well as $S_\lambda^{(1)}(x_p P_p, x_p'P_p)$, which is a three-parton scattering kernel containing an additional gluon from the polarized proton. We should appreciate the appearance of the $k_{1\perp}$-derivative as a consequence of performing the computation up to twist-3 but also due to the connection of $\pd S^{(0)}(k_1)/\pd k_{1\perp}$ with $S^{(1)}_\lambda(x_p P_p,x_p' P_p)$ through the Ward identity for the gluon from the proton \cite{Ratcliffe:1985mp,Benic:2019zvg}.

We point out here that the computation in \cite{Kovchegov:2012ga} is on the parton level, taking transversely polarized spinors $u(p,S_\perp)$ for the initial quark. Thanks to the decomposition $u(p,S_\perp)\bar{u}(p,S_\perp) = (\slashed{p} + m)(1 + \gamma_5 \slashed{S}_\perp)/2$ \cite{Leader:2011vwq} only the $\gamma_5 \slashed{S}_\perp$ Dirac structure is relevant for the polarized quark in the current context (the remaining $S_\perp$-dependent term eventually gets interpreted as the transversity PDF, but this does not contribute in what follows). Thus the computation in \cite{Kovchegov:2012ga} clearly corresponds to the term in \eqref{eq:wtw3} that is proportional to $g_T(x_p)$, where one naturally replaces the quark mass $m$ by the nucleon mass $M_N$.

The above introduced distributions satisfy the QCD equation of motion identity \cite{Eguchi:2006qz,Kanazawa:2015ajw}
\be
x g_T(x) = g^{(1)}_{1T}(x) - \frac{1}{2}\int \rmd x' \frac{G_F(x,x') + \tilde{G}_F(x,x')}{x - x'} ~.
\label{eq:eomr}
\ee
The $g_T(x)$ distribution satisfies another important relation connecting it to the twist-2 helicity PDF $\Delta q(x)$ \cite{Eguchi:2006qz,Kanazawa:2015ajw} thus revealing that $g_T(x)$ itself has a twist-2 piece
\be
g_T(x) = \int_{x}^1 \frac{\rmd x'}{x'} \Delta q(x') + ({\rm genuine \,\, twist-3})\,.
\ee
The remainder is given in terms of the ETQS functions, see \cite{Eguchi:2006qz,Kanazawa:2015ajw} for the explicit expression. Our computation will be based on the WW approximation, that is, taking into account only the $g_T(x_p)$ and the $g_{1T}^{(1)}(x_p)$ contributions to $\rmd \Delta\sigma$ from \eqref{eq:wtw3}. Furthermore, in the WW approximation $g_{1T}^{(1)}(x)$ is fixed through \eqref{eq:eomr} as $g^{(1)}_{1T}(x) \simeq x g_T(x)$ and so \eqref{eq:wtw3} takes the following compact form
\be
E_h\frac{\rmd \Delta\sigma}{\rmd^3 P_h} \simeq  \frac{1}{2(2\pi)^3}\frac{M_N}{2}\int\frac{\rmd z_h}{z_h^2}D(z_h)\int\rmd x_p x_p g_T(x_p)\left(S_\perp^\lambda\frac{\partial}{\partial k_{1\perp}^\lambda}{\rm tr}[\gamma_5 \slashed{k}_1 S^{(0)}(k_1)]\right)_{k_1 = p_1}\,,
\label{eq:wmain}
\ee
that we will take as the starting point of our explicit computations below.

The analogous expression for the gluon initiated channel is adapted from (17) and (25) in \cite{Hatta:2013wsa} (see also (32) in \cite{Benic:2021gya}) to read
\be
\begin{split}
E_h\frac{\rmd \Delta\sigma}{\rmd^3 P_h} & = \frac{1}{2(2\pi)^3}\int\frac{\rmd z_h}{z_h^2} D(z_h)\Bigg[\rmi M_N \int \rmd x_p \calG_{3T}(x_p) \frac{1}{p_1^+}\epsilon^{n\alpha\beta S_\perp} S^{(0)\alpha' \beta'}(p_1)\omega_{\alpha' \alpha}\omega_{\beta'\beta}\\
& - \rmi M_N \int \frac{\rmd x_p}{x_p^2} \tilde{g}(x_p)\left(g_\perp^{\beta\lambda} \epsilon^{\alpha \bar{n}n S_\perp} - g_\perp^{\alpha\lambda}\epsilon^{\beta \bar{n}n S_\perp}\right)\left(\frac{\pd S_{\alpha\beta}^{(0)}(k_1)}{\pd k_1^\lambda}\right)_{k_1 = p_1}\\
& - \frac{1}{2} \int \frac{\rmd x_p \rmd x_p'}{x_p x_p'} M^{\alpha\beta\gamma}_F(x_p,x_p')\frac{S^{(1)\alpha'\beta'\gamma'}(x_p P_p,x_p'P_p)}{x_p' - x_p}\omega_{\alpha' \alpha}\omega_{\beta'\beta}\omega_{\gamma'\gamma}\Bigg]\,,
\end{split}
\label{eq:wqqbar}
\ee
where $\omega_{\alpha\beta} = g_{\alpha\beta} - \bar{n}_\alpha n_\beta$. In the first line we have the intrinsic contribution with $\calG_{3T}(x_p)$ being the gluonic counterpart of $g_T(x_p)$. In the second line $\tilde{g}(x_p)$ is the gluonic kinematical function, see \cite{Hatta:2012jm,Koike:2019zxc} for the definition, and $M^{\alpha\beta\gamma}_F(x_p,x_p')$ is the three-gluon correlator. In the WW approximation $\calG_{3T}(x)$ becomes related to the gluon helicity PDF $\Delta G(x)$ as \cite{Hatta:2012jm}
\be 
\calG_{3T}(x) \simeq \frac{1}{2}  \int_x^1 \frac{\rmd x'}{x'} \Delta G(x')\,,
\ee
while $\tilde{g}(x) \simeq x^2\calG_{3T}(x)$ \cite{Hatta:2012jm}. The WW truncation then amounts to the first two lines of \eqref{eq:wqqbar}.

\subsection{A recap of the leading order inclusive hadron production}

The leading order (LO) amplitude for inclusive hadron production from the $q(k_1) \to q(q)$ channel in the $n\cdot A = A^+ = 0$ gauge is simply given as \cite{Dumitru:2002qt}
\be
\calM = \gamma^+\int_{\xp} \rme^{\rmi(\qp - \khp)\cdot \xp}\left[V(\xp) - 1\right]\,.
\label{eq:mlo}
\ee
Here $V(\xp) = \calP \exp\left[\rmi g \int_{-\infty}^\infty \rmd x^+ A_a^-(x)t^a\right]$ is the fundamental Wilson line with $A_a^-(x)$ being the classical field of the target and we use $\int_{\xp} \equiv \int \rmd^2 \xp$. In \eqref{eq:mlo} we have omitted the overall light-cone delta function, $(2\pi)\delta\left(k_1^+ - q^+\right)$, as well as the initial and final state spinors, thus leaving a matrix in spinor (and color) space. From $\calM$ we obtain the leading order result for $S^{(0)}(k_1)$ as
\be
\begin{split}
S^{(0)}(k_1) &\equiv \frac{1}{2 P_p^+} \frac{1}{N_c}\left\langle \bar{\calM} \slashed{q}\calM \right\rangle (2\pi)\delta\left(k_1^+ - q^+\right)\\
& = (2\pi)\delta\left(k_1^+ - q^+\right) \frac{x_p}{2 k_1^+}\gamma^+ \slashed{q}\gamma^+ \int_{\xp\xp'}\calS(\xp,\xp') \rme^{\rmi(\qp-\khp)\cdot(\xp - \xp')}\,,
\end{split}
\ee
where $1/2P_p^+$ is the flux factor, $1/N_c$ is coming from averaging over the color of the initial state quark and
\be
\calS(\xp,\xp') \equiv \frac{1}{N_c} {\rm tr}\left\langle V(\xp)V^\dag(\xp')\right\rangle\,,
\label{eq:dip}
\ee
is the color averaged dipole distribution. Here and in the following we are suppressing the dependence of the nuclear distributions on the momentum fraction $x_A = k_2^-/P_A^-$, where $k_2$ is the partonic momenta from the nuclei. At the LO we have the momentum conservation $k_1 + k_2 = q$. We readily conclude that ${\rm tr}\left[\gamma_5 \slashed{k}_1 S^{(0)}(k_1)\right] \sim {\rm tr}\left[\gamma_5 \slashed{k}_1 \gamma^+ \slashed{q}\gamma^+\right] \sim \epsilon^{++ q k_1} = 0$ and so \eqref{eq:wmain} vanishes at the LO. 

The analogous expressions in the $g(k_1)\to g(k_g)$ channel are \cite{Dumitru:2005gt}
\be
\calM = (-2 k_1^+) \int_{\xp} \rme^{\rmi(\kgp - \khp)\cdot \xp}\left[U(\xp)-1\right]\,,
\ee
and
\be
\begin{split}
S^{(0)}_{\alpha\beta}(k_1) &= \frac{1}{2P_p^+}\frac{1}{N_c^2 - 1} \langle \calM^\dag \calM\rangle d_{\alpha\beta}(k_g)\\
& = (2\pi)\delta(k_1^+ - k_g^+)(2 k_1^+)^2 d_{\alpha\beta}(k_g) \int_{\xp\xp'} \calS_A(\xp,\xp')\rme^{\rmi(\kgp-\khp)\cdot(\xp - \xp')}\,,
\end{split}
\ee 
where
\be
d^{\alpha\beta}(k) = -g^{\alpha\beta} + \frac{n^\alpha k^\beta + n^\alpha k^\beta}{k^+}\,,
\ee
is the gluon polarization tensor. Note that $d^{\alpha\beta}(p_1) = -g_\perp^{\alpha\beta}$. $U(\xp)$ is the adjoint Wilson line and
\be
\calS_A(\xp,\xp') \equiv \frac{1}{N_c^2 - 1} {\rm tr}\left\langle U(\xp)U^\dag(\xp')\right\rangle\,,
\label{eq:SAdj}
\ee
is the adjoint dipole distribution. The contribution from this channel also vanishes at the LO simply due to the realness of the adjoint Wilson line.

\section{NLO inclusive hadron production in $p^\uparrow A \to hX$: the $q\to qg$ channel}
\label{sec:qg}

At the NLO we have in general the $q\to qg$, $g\to q\bar{q}$ and $g \to gg$ channels. In this Section we compute the $q\to qg$ channel (together with the accompanying virtual contribution), while the $g \to q\bar{q}$ and the $g \to gg$ channels are discussed separately in Secs.~\ref{sec:qqbar} and \ref{sec:ggg}, respectively.

\subsection{$q\to q$: real contribution}
\label{sec:real}

We consider the $q(k_1)\to q(q)g(k_g)$ partonic channel where in the final state a real gluon with momentum $k_g$ gets radiated in addition to the quark with momentum $q$. Here $k_2$ is set by momentum conservation at NLO: $k_1 + k_2 = q + k_g$. We will focus on the case where the quark fragments into a final state hadron and we integrate over the (untagged) gluon phase space to compute the inclusive hadron cross section according to \eqref{eq:wmain}. The main quantity to compute is $S^{(0)}(k_1)$ which takes the following form
\be
\begin{split}
S^{(0)}(k_1) & =  \frac{1}{2P_p^+} \int \frac{\rmd^3 k_g}{(2\pi)^3 2 E_g} \frac{1}{N_c}\left\langle\bar{\calM}^{\mu'}\slashed{q}\calM^{\mu} \right\rangle d_{\mu\mu'}(k_g) (2\pi) \delta\left(k_1^+ - q^+ - k_g^+\right)\,,\\
 & = \frac{q^+}{P_p^+} \int_{\kgp} \frac{1}{N_c}\frac{1}{4q^+k_g^+}\left\langle\bar{\calM}^{\mu'}\slashed{q}\calM^{\mu} \right\rangle d_{\mu\mu'}(k_g)\,,
\end{split}
\label{eq:S0qg}
\ee
where $\int_{\kgp} \equiv \int \frac{\rmd^2 \kgp}{(2\pi)^2}$ and similar for other transverse momenta integrations. Using the quark and gluon propagators in the CGC background \cite{Ayala:1995kg,McLerran:1998nk,Balitsky:2001mr} we can compute the following amplitudes
\be
\begin{split}
\calM_{1}^\mu &= -\rmi g\gamma^{\mu}\frac{\slashed{q}+\slashed{k}_g}{(q+k_g)^2 + \rmi \epsilon}\gamma^+\int_{\xp} \rme^{\rmi\kAp\cdot\xp}t^a[V(\xp)-1],\\
\calM_{2}^\mu &= -\rmi g\gamma^+\frac{\slashed{k}_1-\slashed{k}_g}{(k_1-k_g)^2 + \rmi \epsilon}\gamma^{\mu}\int_{\xp} \rme^{\rmi\kAp\cdot\xp}[V(\xp)-1]t^a,\\
\calM_{3}^\mu &= \rmi g (2k_g^+)\gamma_{\nu}\frac{d^{\nu\mu}(k_1-q)}{(k_1-q)^2 + \rmi \epsilon}\int_{\xp} \rme^{\rmi\kAp\cdot\xp}t^b[U^{ab}(\xp)-\delta^{ab}],\\
\calM_{4}^\mu &= -g (2k_g^+)\int_{\kp}\int_{-\infty}^\infty \frac{\rmd k^-}{2\pi}\gamma^+\frac{\slashed{q}-\slashed{k}}{(q-k)^2 + \rmi \epsilon}\gamma_{\nu}\frac{d^{\nu\mu}(k_1 + k - q)}{(k_1 + k - q)^2 + \rmi \epsilon}\\
& \times\int_{\xp\yp}\rme^{\rmi\kp\cdot\xp}\rme^{\rmi(\kAp-\kp)\cdot\yp}\left[V(\xp)-1 \right]t^b\left[U^{ab}(\yp)-\delta^{ab}\right]\,,
\end{split}
\label{eq:ampqg}
\ee
with the total amplitude $\calM$ given as $\calM^\mu = \sum_{k = 1}^4 \calM_k^\mu$.  We find that the result \eqref{eq:ampqg} agrees with \cite{Jalilian-Marian:2004vhw} except for the overall sign and the adjoint indices in the eikonal gluon vertex that enters $\calM_3^\mu$ and $\calM_4^\mu$.
In the special case when the final quark and gluon are on-shell\footnote{Without loss of generality, the initial quark can be taken as on-shell even at finite $\khp$, that is, $k_1^2 = 0$. Eq.~\eqref{eq:wmain} is not affected due to the $\pd/\pd k_{1\perp}^\lambda$ derivative.}, $\calM^\mu$ takes the following simple form
\be
\calM^\mu = -\rmi g\int_{\kp}\int_{\xp \yp} \rme^{\rmi\kp\cdot \xp} \rme^{\rmi (\kAp - \kp)\cdot \yp} \left[T^\mu_{q} t^a V(\xp) + T^\mu_{qg}(\kp) V(\xp)t^b U^{ab}(\yp)\right]\,,
\label{eq:monsh}
\ee
where
\be
T^\mu_q = \gamma^\mu \frac{\slashed{q} + \slashed{k}_g}{(q+k_g)^2 }\gamma^+\,,
\ee
and
\be
T^\mu_{qg}(\kp) = -\gamma^+ \frac{\slashed{q}-\slashed{k}}{(q-k)^2}\gamma_\nu d^{\nu\mu}(k_1 + k - q)\,.
\label{eq:Tqg2}
\ee
Here, we evaluated the $k^-$ integral in favor of $(k_1 + k - q)^2  +\rmi \epsilon = 0$ so that
\be
(q - k)^2 = - \frac{1}{k_1^+ k_g^+}\left[q^+\khp + k_1^+(\kp - \qp)\right]^2\,.
\ee
Inserting \eqref{eq:ampqg} into \eqref{eq:S0qg} we find
\be
\begin{split}
S^{(0)}(k_1) & = \frac{q^+}{P_p^+}\frac{g^2 C_F}{4q^+ k_g^+} \int_{\kgp\kp\kp'}\int_{\xp\xp'\yp\yp'} \rme^{\rmi\kp \cdot \xp} \rme^{\rmi(\kAp - \kp)\cdot \yp}\\
& \times   \rme^{-\rmi\kp' \cdot \xp'} \rme^{-\rmi(\kAp - \kp')\cdot \yp'} d_{\mu\mu'}(k_g)\Big[\calS(\xp,\xp') \bar{T}_q^{\mu'} \slashed{q} T^\mu_q + \calS_{qqg}(\xp',\xp,\yp')\bar{T}^{\mu'}_{qg}(\kp') \slashed{q} T^\mu_q\\
&   + \calS_{qqg}(\xp',\xp,\yp) \bar{T}_q^{\mu'}\slashed{q}T^\mu_{qg}(\kp) + \calS_{qgqg}(\xp',\yp',\xp,\yp) \bar{T}^{\mu'}_{qg}(\kp')\slashed{q} T^\mu_{qg}(\kp)\Big]\,,
\label{eq:s0main}
\end{split}
\ee
where $\calS(\xp,\xp')$ is the dipole defined in \eqref{eq:dip}, and the additional distributions are given as
\be
\begin{split}
& \calS_{qqg}(\xp',\xp,\yp') \equiv \frac{1}{C_F N_c} \left\langle{\rm tr}\left(V^\dag(\xp') t^b V(\xp) t^a\right)U^{ba}(\yp')\right\rangle\,,\\
& \calS_{qgqg}(\xp',\yp',\xp,\yp) \equiv \frac{1}{C_F N_c} \left\langle{\rm tr}\left( V^\dag(\xp') V(\xp) t^a t^b \right)\left[U^\dag(\yp')U(\yp)\right]^{ba}\right\rangle\,,
\end{split}
\label{eq:corrs}
\ee

 %--- figure ---%
\begin{figure}
  \begin{center}
  \includegraphics[scale = 0.5]{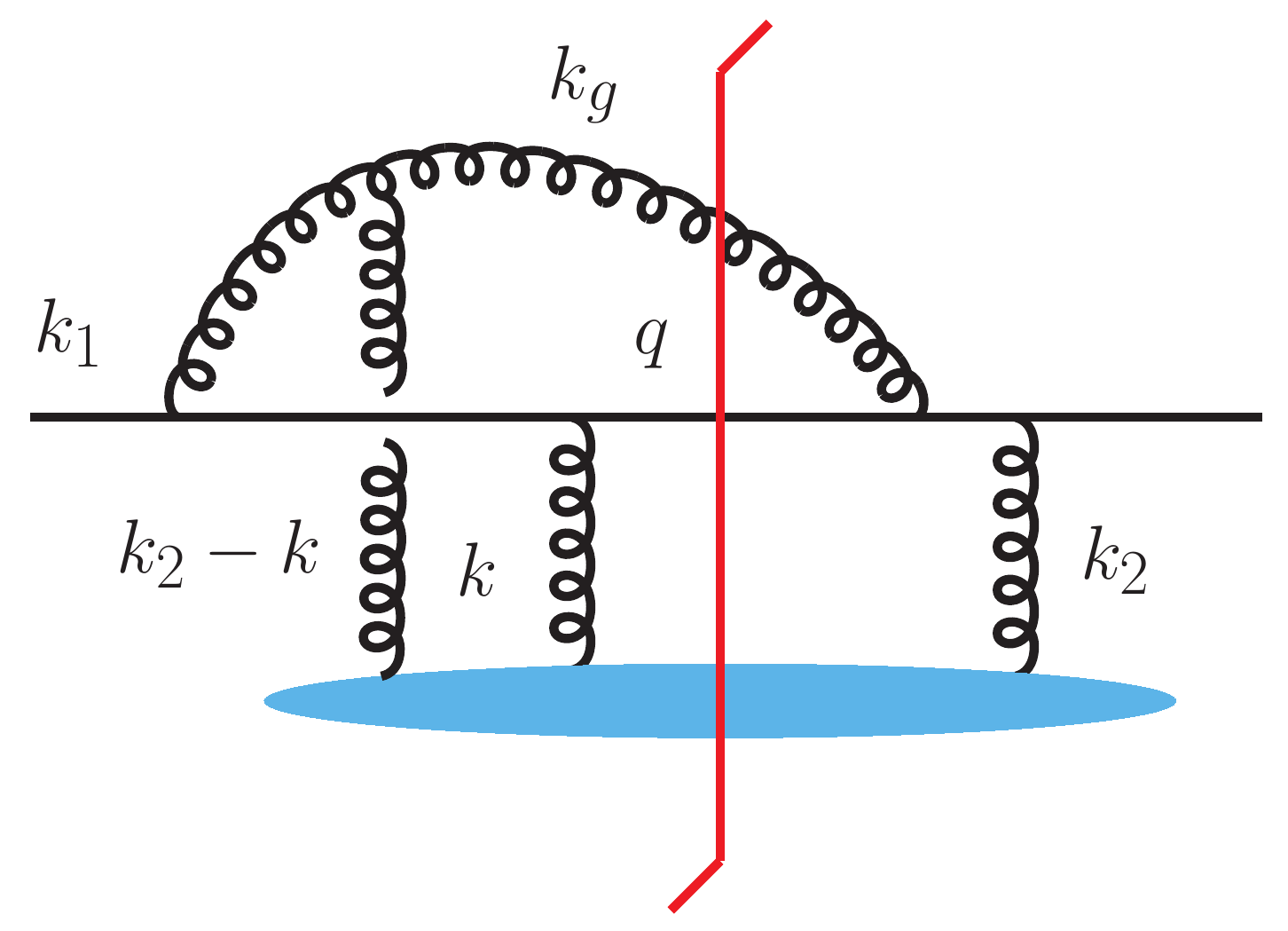}
  \end{center}
  \caption{An interference diagram that determines $S^{(0)}(k_1)$ in the $q \to qg$ channel. The vertical gluons denote Wilson lines arising from multiple scattering on the dense nucleus (represented by a blue blob).}
  \label{fig:qgchan}
\end{figure}
We now show that the first and the fourth term in the square brackets in \eqref{eq:s0main} do not contribute to the polarized cross section when integrated over the gluon momenta. This is intuitively clear as the SSA must come from interferences of different amplitudes, that are given by the second and the third term (c.f., Fig.~\ref{fig:qgchan}, while the first and the fourth term are squares of amplitudes. The analogous structure can also be identified in the computation of \cite{Kovchegov:2012ga}. Consider the first term in \eqref{eq:s0main}, where in the context of \eqref{eq:wmain} we have
\be
{\rm tr}\left[\gamma_5 \slashed{k}_1 \bar{T}_q^{\mu'} \slashed{q} T^\mu_q\right]\,.
\ee
We now apply $C$-parity transformation where $C = \rmi \gamma^0 \gamma^2$ and easily deduce that
\be
\begin{split}
d_{\mu\mu'}(k_g){\rm tr}\left[\gamma_5 \slashed{k}_1 \bar{T}_q^{\mu'} \slashed{q} T^\mu_q\right] & = d_{\mu\mu'}(k_g){\rm tr}\left[\gamma_5 \slashed{k}_1 \bar{T}_q^{\mu'} \slashed{q} T^\mu_q\right]^T = d_{\mu\mu'}(k_g){\rm tr}\left[C (\gamma_5 \slashed{k}_1 \bar{T}_q^{\mu'} \slashed{q} T^\mu_q)^T C^{-1}\right]\\
& = -d_{\mu\mu'}(k_g){\rm tr}\left[\gamma_5 \slashed{k}_1 \bar{T}_q^{\mu'} \slashed{q} T^\mu_q\right]\,,
\end{split}
\ee
since $d_{\mu\mu'}(k_g)$ is symmetric. Therefore, the first term vanishes under the trace by $C$-parity. As for the fourth term we first note that it is in fact independent of $\kgp$. This seems almost trivial as $\kgp$ never enters \eqref{eq:Tqg2}. However, there is a potential $\kgp$ dependence in $d_{\mu\mu'}(k_g)$ given through $d^{--}(k_g) = 2 k_g^-/k_g^+ = \kgp^2/k_g^{+2}$ as well as $d^{-i}(k_g) = k_g^i /k_g^+$. These two terms must couple to $\bar{T}^+_{qg}(\kp')\slashed{q} T^+_{qg}(\kp)$ and $\bar{T}^i_{qg}(\kp')\slashed{q} T^+_{qg}(\kp)$, respectively. 
But, $T^+_{qg}(\kp) \sim d^{\nu+}(k_1 + k - q) = 0$ so there is no  $\kgp$ dependence in the fourth term after all. In order to be able to integrate the fourth term with respect to $\kgp$ we consider the following observation. Defining first
\be
z \equiv \frac{k_g^+}{q^+ + k_g^+}\,, \qquad \bar{z} \equiv 1-z\,,
\ee
as the momentum fraction of the recoiling gluon,
the parton momentum fraction in the projectile is $x_p =q^+/(\bar{z}P_p^+)$. For the target we have
\be
x_A = \frac{q^-}{P_A^-}\left(1 + \frac{k_g^-}{q^-} - \frac{k_1^-}{q^-}\right) = \frac{q^-}{P_A^-}\left[1 + \frac{\bar{z}}{z}\frac{(\khp + \kAp - \qp)^2}{\qp^2} - \bar{z}\frac{\khp^2}{\qp^2}\right]\,.
\ee
Typically in CGC computations, one ignores the dependence of $x_A$ on $\khp$ and $\kAp$, see e.~g. \cite{Chirilli:2012jd}, and uses an approximate formula $x_A \simeq q^-/(zP_A^-)$. The argument is that large values of $\kAp$ should be exponentially suppressed in the cross section due to the nature of the CGC distributions, and the derivative with respect to $\khp$, see \eqref{eq:wmain}, should be $\alpha_S$ suppressed via small-$x$ evolutions \cite{Jalilian-Marian:1997qno,Jalilian-Marian:1997ubg,Iancu:2000hn,Iancu:2001ad}. The former is implicit in \cite{Kovchegov:2012ga} where the computation is based on the initial condition model for the target gluon distributions. With this approximation, the only $\kgp$ dependence is in the phases and this leads to
\be
\int_{\kgp} \rme^{\rmi\kgp(\yp - \yp')} = \delta(\yp - \yp')\,,
\ee
in which case for $\yp' \to \yp$ we have $\calS_{qgqg}(\xp',\yp',\xp,\yp) \to \calS(\xp,\xp')$ which is independent of $\yp$. This allows us to perform the $\yp$ integration, yielding
\be
\int_{\yp} \rme^{\rmi(-\kp + \kp')\cdot \yp} = (2\pi)^2 \delta(\kp - \kp')\,.
\ee
The Dirac trace from the fourth term becomes
${\rm tr}\left[\gamma_5 \slashed{k}_1 \bar{T}_{qg}^{\mu'}(\kp) \slashed{q} T^\mu_{qg}(\kp)\right]$ which vanishes by $C$-parity. Therefore, only the second and the third terms (corresponding to the interference diagram from Fig.~\ref{fig:qgchan}) in \eqref{eq:s0main} are left and we have
\be
\begin{split}
{\rm tr}\left[\gamma_5 \slashed{k}_1 S^{(0)}(k_1)\right] & = \frac{q^+}{P_p^+}g^2 C_F \int_{\kgp\kp\kp'}\int_{\xp\xp'\yp\yp'} \rme^{\rmi\kp \cdot \xp} \rme^{\rmi(\kAp - \kp)\cdot \yp}\rme^{-\rmi\kp' \cdot \xp'} \rme^{-\rmi(\kAp - \kp')\cdot \yp'}\\
& \Big[-\calS_{qqg}(\xp',\xp,\yp')\calH(\kp',\khp) + \calS_{qqg}(\xp',\xp,\yp) \calH(\kp,\khp)\Big]\,,
\label{eq:s0interf}
\end{split}
\ee
where
\be
\begin{split}
&\calH(\kp,\khp) \equiv \frac{1}{4q^+ k_g^+}d_{\mu\mu'}(k_g) {\rm Tr}\left[\gamma_5 \slashed{k}_1 \bar{T}_q^\mu \slashed{q} T_{qg}^{\mu'}(\kp)\right]\,.
\end{split}
\label{eq:hardww}
\ee
In \eqref{eq:s0interf} we have used $\calH^\dag(\kp,\khp) = - \calH(\kp,\khp)$ which is due to the appearance of $\gamma_5$. With the help of the $SU(N_c)$ identity $U^{ab}(\xp) = 2 {\rm tr}\left(t^a V(\xp)t^b V^\dag(\xp)\right)$ and taking the large $N_c$ limit we have
\be
\calS_{qqg}(\xp',\xp,\yp') \simeq \frac{1}{2C_F N_c}\left(N_c^2\calS(\yp',\xp')\calS(\xp,\yp') - \calS(\xp,\xp')\right)\,.
\label{eq:Sqqg}
\ee
The second (dipole) term in \eqref{eq:Sqqg} drops out when combined with the hard factors in \eqref{eq:s0interf}. To see this, note that the $\yp$ and the $\yp'$ integrations in \eqref{eq:s0interf} result in $\delta$-functions that yield $\kp' = \kp = \kAp$. This gives $-\calH(\kAp,\khp) + \calH(\kAp,\khp) = 0$. 

Now we split the dipole into its real and imaginary parts
\be
\calS(\xp,\yp) = \calP(\xp,\yp) + \rmi \calO(\xp,\yp)\,,
\ee
where 
\be
\begin{split}
&\calP(\xp,\yp) \equiv \frac{1}{2}\left(\calS(\xp,\yp) + \calS(\yp,\xp)\right)\,\\
&\calO(\xp,\yp) \equiv \frac{1}{2\rmi}\left(\calS(\xp,\yp) - \calS(\yp,\xp)\right)\,,
\end{split}
\label{eq:defpo}
\ee
are the pomeron and the odderon distributions, respectively. We also replace the primed and unprimed transverse coordinate and momenta labels in the first term in \eqref{eq:s0interf}, namely: $\kp' \leftrightarrow \kp$, $\xp' \leftrightarrow \xp$, $\yp' \leftrightarrow \yp$. By compensating for the reversed sign in the exponentials with $\xp\to -\xp$, and using overall invariance under reflections for the distributions in the unpolarized target, we obtain
\be
\begin{split}
{\rm tr}\left[\gamma_5 \slashed{k}_1 S^{(0)}(k_1)\right] & = \rmi g^2 N_c \frac{q^+}{P_p^+} \int_{\kAp\kp}\int_{\xp\xp'\yp} \rme^{\rmi\kp \cdot (\xp-\yp)} \rme^{-\rmi\kAp\cdot (\xp'-\yp)}\\
&\times \left[\calP(\xp,\yp)\calO(\xp',\yp) - \calO(\xp,\yp)\calP(\xp',\yp)\right]\calH(\kp,\khp)\,.
\end{split}
\label{eq:S0ft}
\ee
Here we have used the following symmetry properties $\calP(\yp,\xp) = \calP(\xp,\yp)$ and $\calO(\yp,\xp) = - \calO(\xp,\yp)$ which follow from \eqref{eq:defpo}. We have also passed from $\kgp$ to $\kAp$ integration. This result clearly demonstrates that the polarized cross section is proportional to the odderon operator.

The Dirac trace is easy to calculate and we find
\be
\calH(\kp,\khp) = 4\rmi(\bar{z}+1)\frac{\vonp \times \vtwp}{\vonp^2 \vtwp^2}\,,
\label{eq:hardnlo}
\ee
where $\vonp \times \vtwp \equiv \epsilon^{-+ v_{1\perp} v_{2\perp}} = v_{1\perp} v_{2\perp} \sin(\phi_1 - \phi_2)$ and
\be
\begin{split}
&\vonp \equiv z\qp - \bar{z}\kgp = \qp - \bar{z}\khp - \bar{z}\kAp\,,\\
&\vtwp \equiv \qp - \bar{z}\khp - \kp\,.
\end{split}
\label{eq:l1l2t}
\ee
Eqs.~\eqref{eq:S0ft} and \eqref{eq:hardnlo} represent the main results of this section.
The vectors $\vonp$, $\vtwp$ reflect the collinear gluon radiations so that when $\vonp \to 0$ ($\vtwp \to 0$) the radiated gluon would be collinear to the final (initial) state quark. Note, however, that in \eqref{eq:hardnlo} both of these limits are completely finite, meaning that the usual collinear divergences one encounters in the NLO computations of an unpolarized cross section for inclusive hadron production, see for example \cite{Chirilli:2012jd}, are absent in this particular computation of the polarized cross section. In addition, when $z \to 0$ ($\bar{z} \to 1$), i.e., when the radiated gluon is collinear to the nucleus ($k_g^+ \to 0$ and so $k_g^- \to \infty$ effectively), the hard factor is also finite. In fact, a close inspection reveals that the resulting cross section is zero in this limit. Namely, when $z\to 1$ there is a symmetry in the hard factor such that by interchanging $\kp \leftrightarrow \kAp$ so that $\vonp \leftrightarrow \vtwp$ the hard factor picks up a sign $\calH \to -\calH$. On the other hand the soft part in \eqref{eq:S0ft} is even under such a transformation and so the overall cross section is zero in this limit. In the case of the NLO unpolarized cross section the $z \to 0$ divergence recovers a part of the small-$x$ evolution of the nuclear wavefunction \cite{Chirilli:2012jd}.

We reflect here also on the computation in \cite{Kovchegov:2012ga} that takes into account only the $g_T(x_p)$ contribution (on the parton level) in \eqref{eq:wmain}. The resulting hard factor associated with $g_T(x)$ is found to be
\be
\calH^{(g_T)}(\kp) = \frac{1}{4q^+ k_g^+}d_{\mu\mu'}(k_g) {\rm tr}\left[\gamma_5 \slashed{S}_\perp \bar{T}_q^\mu \slashed{q} T_{qg}^{\mu'}(\kp)\right]_{\khp = 0} = 4\rmi \bar{z}^2 \frac{\hvonp\times \Sp}{\hvonp^2\hvtwp^2}\,,
\label{eq:hardgt}
\ee
with $\hvonp$ and $\hvtwp$ obtained from $\vonp$ and $\vtwp$ by setting $\khp = 0$, see \eqref{eq:l1l2t}. It is important to observe that, while the final state collinear divergence ($\hvonp\to 0$) is absent, the hard factor has a divergence when the radiated gluon is collinear to the initial state proton ($\hvtwp \to 0$). This divergence is also present in \cite{Kovchegov:2012ga}, as can be seen from their Eq.~(15) by setting the quark mass $m \to 0$\footnote{One should be careful here in first factoring out one power of $m$ in (15) in \cite{Kovchegov:2012ga}, as per the definition of $g_T(x)$.}. In hindsight, this means that the result in \cite{Kovchegov:2012ga} must be incomplete in the sense that the lowest order computation should be free from any divergences. That is, by taking into account also the $g_{1T}^{(1)}(x)$ part of the full cross section \eqref{eq:wtw3}, as per the WW approximation \eqref{eq:wmain}, we indeed find that the initial state collinear divergence is cancelled between the $g_T(x)$ and the $g_{1T}^{(1)}(x)$ parts, resulting in a finite hard factor \eqref{eq:hardnlo}. A similar conclusion was also reached in a collinear framework in SIDIS, see \cite{Benic:2019zvg} and \cite{Benic:2021gya} where the $g_T(x)$ contribution to the cross section contained an initial state collinear divergence, that gets exactly cancelled with the collinear divergence in the $g_{1T}^{(1)}(x)$ part.

\subsection{Proof that the real contribution in the $q\to q$ channel vanishes}
\label{sec:proof}

We now argue that in fact \eqref{eq:S0ft} is exactly zero. Before performing an explicit computation we can appreciate it in an intuitive way as follows. In general for the polarized cross section to be non-zero we need two vectors: the transverse momentum of the final state and the spin so that we can form the familiar cross product $\qp\times \Sp$. In case of \eqref{eq:S0ft}, we have $\qp$, while we can think of $\khp$ as a proxy for the spin, thanks to the derivative $S_\perp^\lambda \pd/\pd k_{1\perp}^\lambda$. However, owing to the particular form of the hard factor \eqref{eq:hardnlo}, the two vectors $\qp$ and $\khp$ enter the cross section only through the linear combination $\qop \equiv \qp - \bar{z} \khp$ (the soft part of the cross section is independent of $\khp$). Thus the final result depends only on a single vector, $\qop$, and therefore must be zero.

To see the above statement explicitly we start by switching to the coordinates
\be
\begin{split}
&\rp = \xp - \yp \,, \qquad \bp  =\frac{\xp + \yp}{2}\,,\\
&\rp' = \yp - \xp' \,, \qquad \bp'  =\frac{\yp + \xp'}{2}\,,
\end{split}
\ee
to obtain
\be
\begin{split}
{\rm tr }\left[\gamma_5 \slashed{k}_1 S^{(0)}(k_1)\right] & = \rmi g^2 N_c \frac{q^+}{P_p^+}\int_{\kAp\kp}\int_{\rp\bp\rp'} \rme^{\rmi\kp \cdot \rp} \rme^{\rmi\kAp\cdot \rp'}\\
&\times \left[\calP(\rp,\bp)\calO(\rp',\bp') - \calO(\rp,\bp)\calP(\rp',\bp')\right]\calH(\kp,\khp)\,.
\end{split}
\label{eq:S0real}
\ee
Note that not all transverse coordinates in \eqref{eq:S0real} are independent: we have the following relation for $\bp'$
\be
\bp' = \bp - \frac{1}{2}(\rp + \rp')\,.
\label{eq:bprime}
\ee
This is an important point because $\calO(\rp,-\bp) = -\calO(\rp,\bp)$ and so an integral over $\bp$ would superficially vanish simply via $\bp \to - \bp$. Next, in order to de-convolve the transverse integrals in \eqref{eq:S0real} we Fourier transform the distributions as
\be
\calP(\rp,\bp) = \int_{\kapp\delp} \rme^{-\rmi\kapp\cdot\rp}\rme^{-\rmi\delp\cdot\bp}\calP(\kapp,\delp)\,,
\ee
and similarly for $\calO(\rp,\bp)$. In terms of the Fourier-transformed distributions,  \eqref{eq:S0real} becomes
\be
\begin{split}
{\rm tr}[\gamma_5 \slashed{k}_1 S^{(0)}(k_1)] &= \rmi g^2 N_c\frac{q^+}{P_p^+}\int_{\kapp\kapp'\delp}\left[\calP(\kapp,\delp)\calO(\kapp',\delp') - \calO(\kapp,\delp)\calP(\kapp',\delp')\right] \calH(\kp,\khp)\,,
\label{eq:S0ft2}
\end{split}
\ee
where $\kp = \kapp + \frac{1}{2}\delp$, $\kAp = \kapp' + \frac{1}{2}\delp$ and $\delp' = - \delp$. 

The key quantity to consider in \eqref{eq:S0ft2} is the integral over the angular variables. While the pomeron carries no angular dependence, the odderon has the following modulation $\calO(\kapp,\delp) \propto (\kapp\cdot \delp)$ (this is simply the momentum space counterpart of the more familiar $\calO(\rp,\bp) \sim (\rp\cdot\bp)$ modulation), see e.~g. \cite{Lappi:2016gqe,Dong:2018wsp,Boussarie:2019vmk}. Focusing on the first part in \eqref{eq:S0ft2} we start from the following expression, 
\be
\int_0^{2\pi} \frac{\rmd \phi_\Delta}{2\pi}\int_0^{2\pi} \frac{\rmd \phi_\kappa}{2\pi}\int_0^{2\pi} \frac{\rmd \phi_{\kappa'}}{2\pi} (\kapp'\cdot \delp') \frac{(\vonp\times \vtwp)}{\vonp^2 \vtwp^2}\,.
\ee
Introducing $\donp \equiv \qop - \bar{z}\delp/2$, $\dtwp \equiv \qop - \delp/2$,
we have
\be
\vonp \times \vtwp = \donp \times \dtwp - \donp \times \kapp - \bar{z} \kapp' \times \dtwp + \bar{z} \kapp \times \kapp'\,,
\label{eq:psa1a2}
\ee
and
\be
\begin{split}
&\vonp^2 = \donp^2 + \bar{z}^2 \kapp'^2 - 2 \bar{z}\kappa'_\perp \delta_{1\perp} \cos(\phi_{\kappa'} - \phi_{\delta_1})\,,\\
&\vtwp^2 = \dtwp^2 + \kapp^2 - 2\kappa_\perp \delta_{2\perp} \cos(\phi_\kappa - \phi_{\delta_2})\,,
\end{split}
\ee
Now we compute the integrals over $\phi_\kappa$ and $\phi_{\kappa'}$. From the first term in \eqref{eq:psa1a2} we obtain
\be
\int_0^{2\pi} \frac{\rmd \phi_\kappa}{2\pi}\int_0^{2\pi} \frac{\rmd \phi_{\kappa'}}{2\pi} (\kapp'\cdot \delp')\frac{(\donp\times \dtwp)}{\vonp^2 \vtwp^2} = -\frac{1}{4}\frac{z}{\bar{z}}(\donp \cdot \delp)\frac{(\qop \times \delp)}{|\dtwp^2 -\kapp^2|}\left(1 - \frac{\donp^2 + \bar{z}^2 \kapp'^2}{|\donp^2 - \bar{z} ^2\kapp'^2|}\right)\,.
\ee
where we used $\donp\times \dtwp = - z (\qop \times \delp)/2$. Inserting now the definition of $\donp$ we will in general have an expression of the type
\be
\int_0^{2\pi} \frac{\rmd \phi_\Delta}{2\pi} \sin(\phi_{q_1}- \phi_\Delta) f\left(\cos(\phi_{q_1} - \phi_{\Delta})\right) = \int_{\phi_{q_1}}^{\phi_{q_1} - 2\pi} \frac{\rmd \phi}{2\pi} \sin\phi f(\cos\phi)\,,
\ee
but this is simply zero as
\be
\int_{\phi_1}^{\phi_1 - 2\pi} \rmd \phi \sin\phi f(\cos\phi) = -\int_{\phi_1}^{\phi_1 - 2\pi} \rmd (\cos\phi) f(\cos\phi) = F(\cos\phi)|_{\phi_1}^{\phi_1 - 2\pi}  = 0\,,
\label{eq:zero}
\ee
where $F(\cos\phi)$ is a primitive function of $f(\cos\phi)$.
By a completely analogous computation we can show that each of the remaining three pieces in \eqref{eq:psa1a2} is also zero and thus conclude that the complete real contribution in the $q\to q$ channel vanishes.

\subsection{$q\to q$: virtual contribution}
\label{sec:virt}

For the virtual correction to the $q\to q$ channel we have the following amplitude
\be
\begin{split}
\mathcal{M} &= g^2\int\frac{\rmd k_g^+}{(2\pi)}\int_{\kgp\kp}\int_{\xp\yp} \rme^{\rmi\kp\cdot\xp} \rme^{\rmi(\qp-\khp-\kp)\cdot\yp}\\
&\times\Big[t^at^aV(\xp)\calT_{q,1}+V(\xp)t^at^a \calT_{q,2}+t^aV(\xp)t^bU^{ab}(\yp)\calT_{qg}\Big],
\end{split}
\label{eq:mvirt}
\ee
where
\be
\begin{split}
\calT_{q,1}&= \rmi \int\frac{\rmd k_g^-}{(2\pi)}\gamma^{\mu}\frac{\slashed{q}-\slashed{k}_g}{(q-k_g)^2 + \rmi \epsilon}\gamma^{\nu}\frac{\slashed{q}}{q^2 + \rmi \epsilon}\gamma^+\frac{d_{\mu\nu}(k_g)}{k_g^2 + \rmi \epsilon}\,,\\
\calT_{q,2}&=\rmi \int\frac{\rmd k_g^-}{(2\pi)} \gamma^+\frac{\slashed{k}_1}{k_1^2 + \rmi \epsilon}\gamma^{\mu}\frac{\slashed{k}_1-\slashed{k}_g}{(k_1-k_g)^2 + \rmi\epsilon}\gamma^{\nu}\frac{d_{\mu\nu}(k_g)}{k_g^2 + \rmi \epsilon}\,\\
\calT_{qg}&=\rmi \int\frac{\rmd k_g^-}{(2\pi)}\gamma^\mu\frac{\slashed{q}-\slashed{k}_g}{(q-k_g)^2 + \rmi\epsilon}\gamma^+\frac{\slashed{q}-\slashed{k}-\slashed{k}_g}{(q - k - k_g)^2 + \rmi\epsilon}\gamma_{\rho}d^{\rho\nu}(k+k_1+k_g-q)\frac{d_{\mu\nu}(k_g)}{k_g^2 + \rmi \epsilon} \,.
\end{split}
\ee
Above, in the first line of $\calT_{qg}$ we have evaluated the $k^-$ integration in favor of the singularity at $(k_1 + k + k_g - q)^2 + \rmi \epsilon = 0$ so that
\be
(q-k-k_g)^2 = -\frac{1}{k_1^+ k_g^+}\left[k_1^+ (\kp + \khp + \kgp - \qp) - k_g^+ \khp\right]^2\,.
\ee
 %--- figure ---%
\begin{figure}
  \begin{center}
  \includegraphics[scale = 0.5]{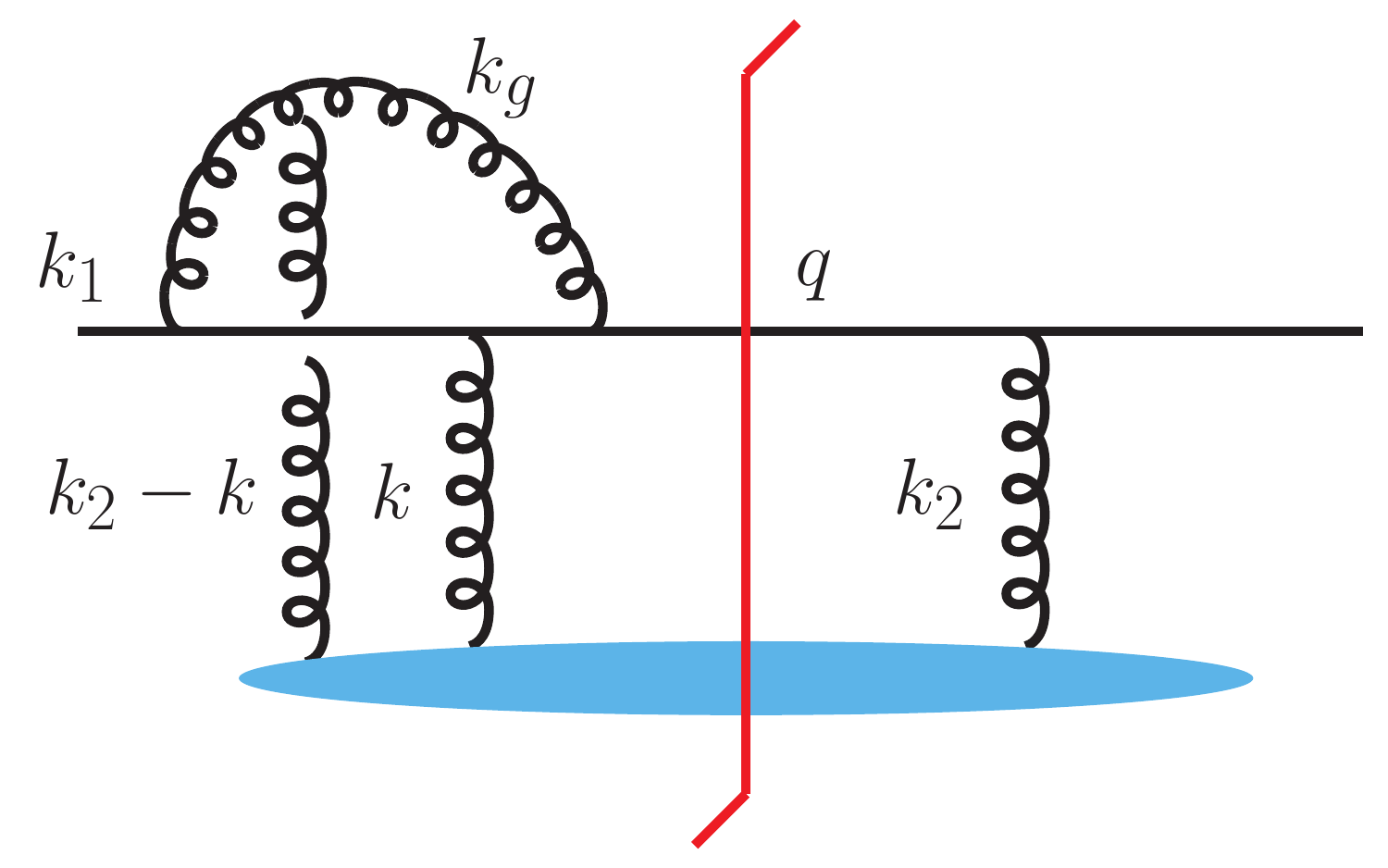}
  \end{center}
  \caption{An interference diagram that determines $S^{(0)}(k_1)$ in the virtual correction to the $q \to q$ channel. The vertical gluons denote Wilson lines arising from multiple scattering on the dense nucleus.}
  \label{fig:qgvirt}
\end{figure}
To get the NLO virtual contribution to $S^{(0)}(k_1)$ we combine the virtual amplitude \eqref{eq:mvirt} with the LO amplitude \eqref{eq:mlo} and find
\be
\begin{split}
S^{(0)}(k_1)&=(2\pi)\delta(k_1^+ - q^+)C_F g^2\frac{1}{2P_p^+}\int\frac{\rmd k_g^+}{(2\pi)}\int_{\kgp\kp\kp'}\int_{\xp\xp'\yp\yp'}\rme^{\rmi\kp\cdot\xp}\rme^{\rmi(\kAp-\kp)\cdot\yp}\rme^{-\rmi\kp'\cdot\xp'}\rme^{-\rmi(\kAp-\kp')\cdot\yp'}\\
&\times\Big[\calS_q(\xp',\xp)\gamma^+\slashed{q}\calT_{q}+\calS_q(\xp',\xp)\bar{\calT}_{q}\slashed{q}\gamma^+\\
&+\calS_{qqg}(\xp',\xp,\yp)\gamma^+\slashed{q}\calT_{qg}(\kp)+\calS_{qqg}(\xp',\xp,\yp')\bar{\calT}_{qg}(\kp')\slashed{q}\gamma^+\Big]\,,
\end{split}
\label{eq:S0virt}
\ee
where now $\kAp \equiv \qp - \khp$ and $\calT_{q} \equiv \calT_{q,1} + \calT_{q,2}$.
Analogous to the case of real production, the terms in \eqref{eq:S0virt} that are proportional to the dipole operator will not contribute as a consequence of $C$-parity. This includes the entirety of the second line and the dipole pieces of the third line according to \eqref{eq:Sqqg}. Repeating further the steps of the calculation used for real production we find
\be
\begin{split}
{\rm tr}\left[\gamma_5 \slashed{k}_1 S^{(0)}(k_1)\right] &= \rmi (2\pi)\delta(k_1^+ - q^+)N_c g^2 \frac{q^+}{P_p^+}\int\frac{\rmd k_g^+}{(2\pi)}\int_{\kgp\kp}\int_{\xp\xp'\yp}\rme^{\rmi\kp\cdot(\xp - \yp)}\rme^{-\rmi\kAp\cdot(\xp' - \yp)}\\
&\times\left[\calP(\xp,\yp)\calO(\xp',\yp) - \calO(\xp,\yp)\calP(\xp',\yp)\right]\calH(\kp,\khp)\,.
\end{split}
\ee
See Fig.~\ref{fig:qgvirt} for the diagram corresponding to this remaining contribution. Here now $\calH(\kp,\khp)$ is
\be
\calH(\kp,\khp) \equiv \frac{1}{2q^+}{\rm tr}\left[\gamma_5 \slashed{k}_1 \gamma^+ \slashed{q}\calT_{qg}(\kp)\right] = -4\rmi (\bar{y} + 1) \frac{\vonp\times\vtwp}{\vonp^2 \vtwp^2}\,,
\label{eq:hardvirt}
\ee
where we have introduced  $y \equiv k_g^+ /k_1^+ = k_g^+ /q^+ = z/\bar{z}$ and $\vonp$ ($\vtwp$) are associated with final (initial) state collinear configurations explicitly given as $\vonp \equiv y \qp - \kgp$, $\vtwp \equiv \kp + \bar{y}\khp + \kgp - \qp$. To compute \eqref{eq:hardvirt} we have evaluated the $k_g^-$ integral in $\calT_{qg}(\kp)$ in favor of the singularity $k_g^2 + \rmi \epsilon = 0$.
Proceeding with the $\kgp$ loop integral we pass from the variable $\kgp$ to $\vtwp$ and write
\be
\int_{\kgp} \frac{\vonp\times \vtwp}{\vonp^2 \vtwp^2} = -\int_{\vtwp} \frac{\vp\times \vtwp}{(\vp + \vtwp)^2 \vtwp^2}\,,
\ee
where $\vp \equiv \bar{y}\qp - y \khp - \kp$. But this contains an angular integral that is precisely of the form \eqref{eq:zero} and therefore vanishes.

\subsection{$q \to g$}

In this case we only have the real diagram with gluon fragmenting into a final state hadron. The expression for $S^{(0)}(k_1)$ takes the same form as \eqref{eq:s0main}, with the only difference being that now we are integrating over $\qp$ (the momenta of the untagged quark) instead of over $\kgp$,
\be
\begin{split}
S^{(0)}(k_1) & = \frac{k_g^+}{P_p^+}\frac{g^2 C_F}{(2q^+)(2 k_g^+)} \int_{\qp\kp\kp'}\int_{\xp\xp'\yp,\yp'} \rme^{\rmi\kp \cdot \xp} \rme^{\rmi(\kAp - \kp)\cdot \yp}\\
& \times   \rme^{-\rmi\kp' \cdot \xp'} \rme^{-\rmi(\kAp - \kp')\cdot \yp'} d_{\mu\mu'}(k_g)\Big[\calS(\xp,\xp') \bar{T}_q^{\mu'} \slashed{q} T^\mu_q + \calS_{qqg}(\xp',\xp,\yp')\bar{T}^{\mu'}_{qg}(\kp') \slashed{q} T^\mu_q\\
&   + \calS_{qqg}(\xp',\xp,\yp) \bar{T}_q^{\mu'}\slashed{q}T^\mu_{qg}(\kp) + \calS_{qgqg}(\xp',\yp',\xp,\yp) \bar{T}^{\mu'}_{qg}(\kp')\slashed{q} T^\mu_{qg}(\kp)\Big]\,.
\label{eq:s0main2}
\end{split}
\ee
The first term again vanishes due to $C$-parity. To show that last term vanishes, we first need to make following replacements: $\kp-\qp\to\kp$ and $\kp'-\qp\to\kp'$ for the $\kp$ and $\kp'$ integrals which makes $T_{gq}(\kp)$ independent of $\qp$. Additionally, since $\slashed{q}$ is sandwiched between two $\gamma^+$ matrices it does not give a $\qp$ contribution. Therefore, the respective hard factor does not depend on $\qp$, and the only $\qp$ dependence in the last term appears in the exponential. From here we take the analogous steps as in the $q\to q$ channel. First performing the $\qp$ integration
\be
\int_{\qp}\rme^{\rmi\qp(\xp-\xp')}=\delta^{(2)}(\xp-\xp')\,,
\ee
$S_{qgqg}(\xp',\yp',\xp,\yp)$ collapses to an adjoint dipole, see \eqref{eq:corrs} when $\xp' = \xp$ which is in addition independent of $\xp$. This allows us to perform the $\xp$ integral to conclude that $\kp' = \kp$. With this we can utilize $C$-parity to find the hard factor from the last term vanishes. For the remaining interference terms the trace is given in \eqref{eq:hardnlo}. With the vectors $\vonp$ and $\vtwp$ suitably re-written in the form $\vonp  = (z\khp - \kgp) + \kAp$ and $\vtwp  = (z\khp - \kgp) + \kAp - \kp$, where we see that a unique combination $z\khp - \kgp$ appears. Thus the steps to show that the cross section also vanishes in the $q\to g$ channel are from this point completely analogous to those for the $q\to q$ channel, see Sec.~\ref{sec:proof}.
Together with the result from Sec.~\ref{sec:proof} and Sec.~\ref{sec:virt} this completes the statement that in the $q\to qg$ channel the the odderon mechanism in the WW approximation does not contribute to SSA at NLO.

\section{The $g\to q\bar{q}$ channel}
\label{sec:qqbar}

In this channel we label the momenta as $g(k_1) \to q(q)\bar{q}(p)$. The NLO amplitude can be written as
\be
\calM^\alpha = (+g)\int_{\kp}\int_{\xp\yp}\rme^{\rmi \kp \cdot \xp} \rme^{\rmi (\kAp - \kp)\cdot \yp}\left[T_g^\alpha t^b U^{ba}(\xp) + T_{q\bar{q}}^\alpha(\kp)V(\xp)t^a V^\dag(\yp)\right]\,,
\label{eq:ampqqbar}
\ee
where
\be
T_g^\alpha \equiv 2 k_1^+  \gamma_\rho \frac{d^{\rho\alpha}(q+p)}{(q+p)^2}\,
\ee
and
\be
T_{q\bar{q}}^\alpha(\kp) \equiv \frac{1}{2p^+}\gamma^+ \frac{\slashed{q}-\slashed{k}}{(q-k)^2}\gamma^\alpha (\slashed{q} - \slashed{k}-\slashed{k}_1)\gamma^+\,.
\ee
In the above expressions, similar to the discussion in Sec.~\ref{sec:real}, $k^+$ is obtained by picking up the pole from the condition $(q-k - k_1)^2 + \rmi \epsilon = 0$. The above results can be shown to agree with the corresponding amplitude in Eq.~(38) in \cite{Blaizot:2004wv} used for unpolarized $pA$ collisions after taking the collinear limit for the gluon from the proton. Using \eqref{eq:ampqqbar} we calculate $S^{(0)}_{\alpha\beta}(k_1)$ as
\be
\begin{split}
S^{(0)}_{\alpha\beta}(k_1) &= \frac{1}{2 P_p^+} \int \frac{\rmd^3 p}{(2\pi)^3 2 E_p} \frac{1}{N_c^2 - 1}{\rm Tr}\langle \bar{\calM}_\alpha \slashed{q}\calM_\beta \slashed{p}\rangle\\
&= \frac{q^+}{P_p^+}\frac{g^2 T_R}{(2q^+)(2p^+)}\int_{\pp\kp\kp'}\int_{\xp\xp'\yp\yp'}\rme^{\rmi \kp\cdot\xp}\rme^{\rmi(\kAp - \kp)\cdot \yp}\rme^{-\rmi \kp'\cdot\xp'}\rme^{-\rmi(\kAp - \kp')\cdot \yp'}\\
&\times\Big\{\calS_A(\xp,\xp') {\rm tr}\left[\bar{T}'_{g,\alpha} \slashed{q}T_{g,\beta}\slashed{p}\right] + \calS_{qqg}(\xp',\yp',\xp) {\rm tr}\left[\bar{T}'_{q\bar{q},\alpha}(\kp') \slashed{q}T_{g,\beta}\slashed{p}\right]\\
&+\calS_{qqg}(\yp,\xp,\xp') {\rm tr}\left[\bar{T}'_{g,\alpha} \slashed{q}T_{q\bar{q},\beta}(\kp)\slashed{p}\right] + \calS_{qqqq}(\xp',\xp,\yp',\yp) {\rm tr}\left[\bar{T}'_{q\bar{q},\alpha}(\kp') \slashed{q}T_{q\bar{q},\beta}(\kp)\slashed{p}\right]\Big\}\,,
\end{split}
\label{eq:S0qqbar}
\ee
where $T_R = 1/2$. Here $\calS_A$ and $\calS_{qqg}$ are defined in \eqref{eq:SAdj} and the first line of \eqref{eq:corrs} respectively, while
\be
\calS_{qqqq}(\xp',\xp,\yp',\yp) \equiv \frac{1}{C_F N_c}{\rm tr}\left\langle V^\dag(\xp)V(\xp)t^aV^\dag(\yp)V(\yp')t^a\right\rangle\,,
\ee
is an additional gluon distribution of the target.
 %--- figure ---%
\begin{figure}
  \begin{center}
  \includegraphics[scale = 0.5]{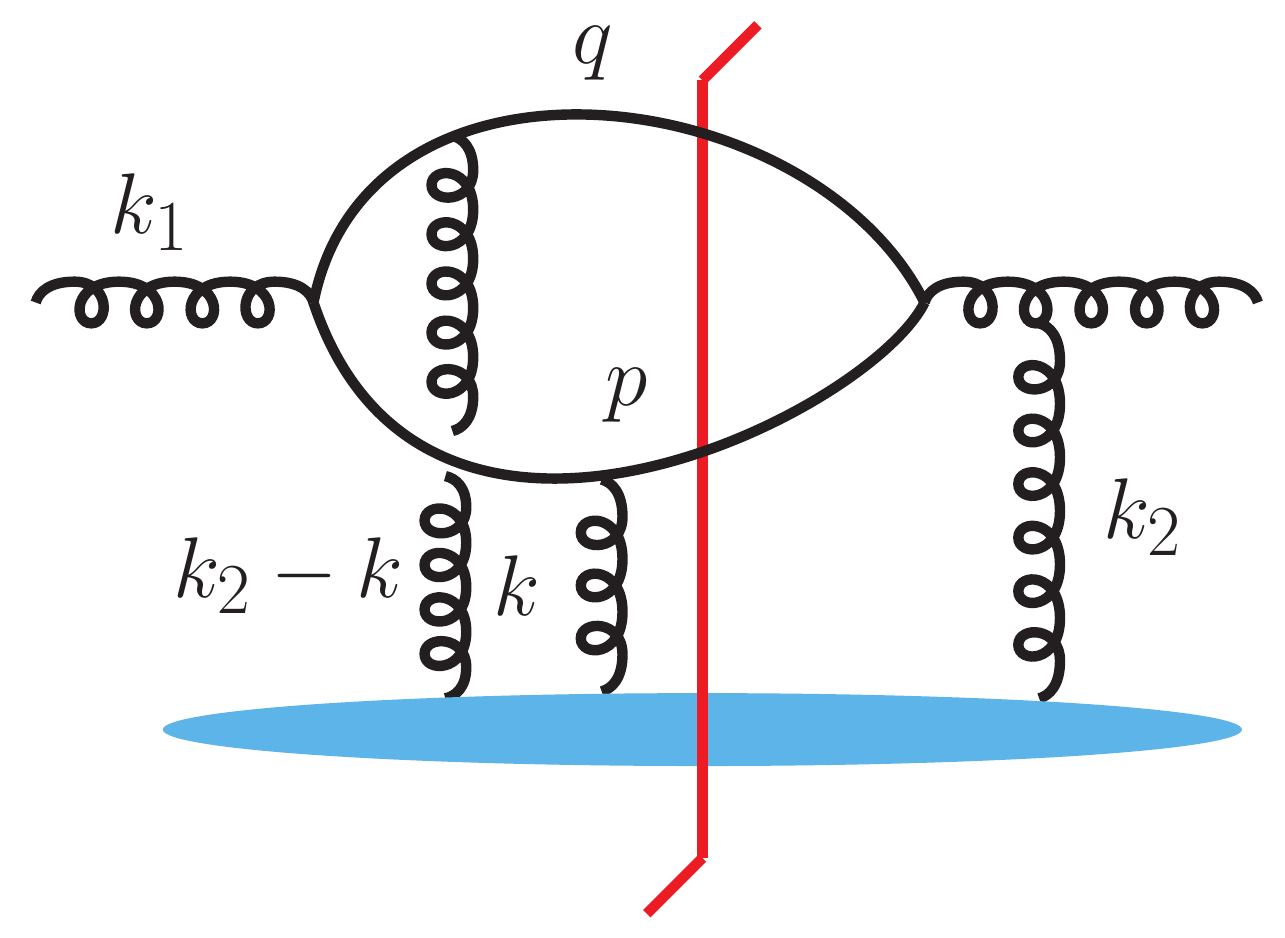}
  \end{center}
  \caption{An interference diagram that determines $S_{\alpha\beta}^{(0)}(k_1)$ in the $g \to q\bar{q}$ channel. The vertical gluons denote Wilson lines arising from multiple scattering on the dense nucleus.}
  \label{fig:qqbar}
\end{figure}
Similar to the findings in Sec.~\ref{sec:real}, the first and the fourth term in \eqref{eq:S0qqbar} vanish. For the first term this can be argued from $C$-parity on the Dirac trace (another way is simply from the fact that the adjoint dipole $\calS_A(\xp,\xp')$ is real). For the fourth term the key point is that the hard factor does not depend on $\pp$ ($\slashed{p}$ is sandwiched between $\gamma^+$ and so the $\pp$ dependence drops out). Then, the fourth term does not contribute by the same steps used in Sec.~\ref{sec:real}. This leaves the interference term in \eqref{eq:S0qqbar} that is represented graphically in Fig.~\ref{fig:qqbar}.  According to the WW truncation of \eqref{eq:wqqbar} these require the evaluation of the following hard factors
\be
\begin{split}
&\calH^{(\calG_{3T})}(\kp) \equiv \frac{1}{4 q^+ p^+}\frac{1}{p_1^+} \epsilon^{+\alpha\beta S_\perp}\omega_{\alpha'\alpha}\omega_{\beta'\beta}{\rm tr}\left[\bar{T}^{\alpha'}_g \slashed{q} T^{\beta'}_{q\bar{q}}(\kp)\slashed{p}\right]\,,\\
& \calH^{(\tilde{g}),\lambda}(\kp,\khp) \equiv \frac{1}{4 q^+ p^+} \left(g_\perp^{\beta\lambda} \epsilon^{\alpha -+ S_\perp} - g_\perp^{\alpha\lambda}\epsilon^{\beta -+ S_\perp}\right){\rm tr}\left[\bar{T}_{g,\alpha} \slashed{q} T_{q\bar{q},\beta}(\kp)\slashed{p}\right]\,.
\end{split}
\ee
We find
\be
\calH^{(\calG_{3T})}(\kp) = 4z\bar{z}(z-\bar{z})\frac{\hvonp\times \Sp}{\hvonp^2 \hvtwp^2}\,,
\label{eq:HG3T}
\ee
\be
\begin{split}
\calH^{(\tilde{g}),\lambda}(\kp,\khp) = \frac{4}{\vonp^2 \vtwp^2}\Big\{ & -\left[(z^2 + \bar{z}^2)(\vtwp\times \Sp) + z\bar{z}(z- \bar{z})(\khp\times \Sp)\right] v_{1\perp}^\lambda\\
& + (z^2 + \bar{z}^2)(\vonp\times \Sp) v_{2\perp}^\lambda\\
& + z\bar{z}(z- \bar{z})\left(\vonp\times \Sp\right)k_{1\perp}^\lambda \Big\}\,,
\end{split}
\ee
where now $\vonp \equiv z\qp - \bar{z}\pp = \qp - \bar{z}\khp - \bar{z}\kAp$, $\vtwp \equiv \qp - \bar{z}\khp - \kp$, with $z \equiv p^+/k_1^+$ the momentum fraction of the recoiling antiquark. By $\hvonp$ ($\hvtwp$) in \eqref{eq:HG3T} we again denote $\vonp$ ($\vtwp$) at $\khp = 0$. According to the WW truncation of the polarized cross section we also need to take a derivative of $\calH^{(\tilde{g}),\lambda}(\kp,\khp)$ with respect to $k_{1\lambda}$ (and sum over $\lambda$), c.f., 2nd line in \eqref{eq:wqqbar}. After this, we could proceed with the angular integrals as in Sec.~\ref{sec:proof}, but this time the expressions would involve $\Sp$. A simpler way to proceed is to first combine the 1st and the 2nd line in \eqref{eq:wqqbar} leading to
\be
\begin{split}
\calH^{(\calG_{3T})}(\kp) & - \left[\frac{\pd}{\pd k_1^\lambda}\calH^{(\tilde{g}),\lambda}(\kp,\khp)\right]_{k_1 = p_1} = -\frac{4\bar{z}(z^2 + \bar{z}^2)}{\hvonp^4 \hvtwp^4}\\
&\times\left[(\hvonp^2 \hvtwp^2 + 2(\hvonp\cdot\hvtwp)\vtwp^2)(\hvonp\times \Sp) - (\hvonp^2 \hvtwp^2 + 2(\hvonp\cdot\hvtwp)\hvonp^2)(\hvtwp\times \Sp)\right]\,.
\end{split}
\label{eq:hardqqbar}
\ee
After some inspection, this can be also re-written in a more convenient form as
\be
\calH^{(\calG_{3T})}(\kp) - \left[\frac{\pd}{\pd k_1^\lambda}\calH^{(\tilde{g}),\lambda}(\kp,\khp)\right]_{k_1 = p_1} = \left[S_{\perp}^\lambda \frac{\pd}{\pd k_{1\perp}^\lambda}\left(4(z^2 + \bar{z}^2)\frac{\vonp\times \vtwp}{\vonp^2 \vtwp^2}\right)\right]_{k_1 = p_1}\,,
\label{eq:hardqqbar2}
\ee
where $S_{\perp}^\lambda$ is now factored out and the effective hard factor inside the brackets in \eqref{eq:hardqqbar2} now has finite $\khp$ through $\vonp$ and $\vtwp$. To prove the equivalence of \eqref{eq:hardqqbar} and \eqref{eq:hardqqbar2} we have used the Schouten identity $\hvonp (\hvtwp \times \Sp) + \hvtwp (\Sp \times \hvonp) + \Sp (\hvonp \times \hvtwp) = 0$. Eq.~\eqref{eq:hardqqbar2} reveals that the general structure of the $g \to q$ hard factor is the same as in the $q\to q$ channel, see \eqref{eq:hardnlo}. Therefore, by following the same logic as in Sec.~\ref{sec:proof} we conclude that the corresponding polarized cross section in the $g \to q\bar{q}$ channel also vanishes.

\section{The $g\to gg$ channel}
\label{sec:ggg}

 %--- figure ---%
\begin{figure}
  \begin{center}
  \includegraphics[scale = 0.5]{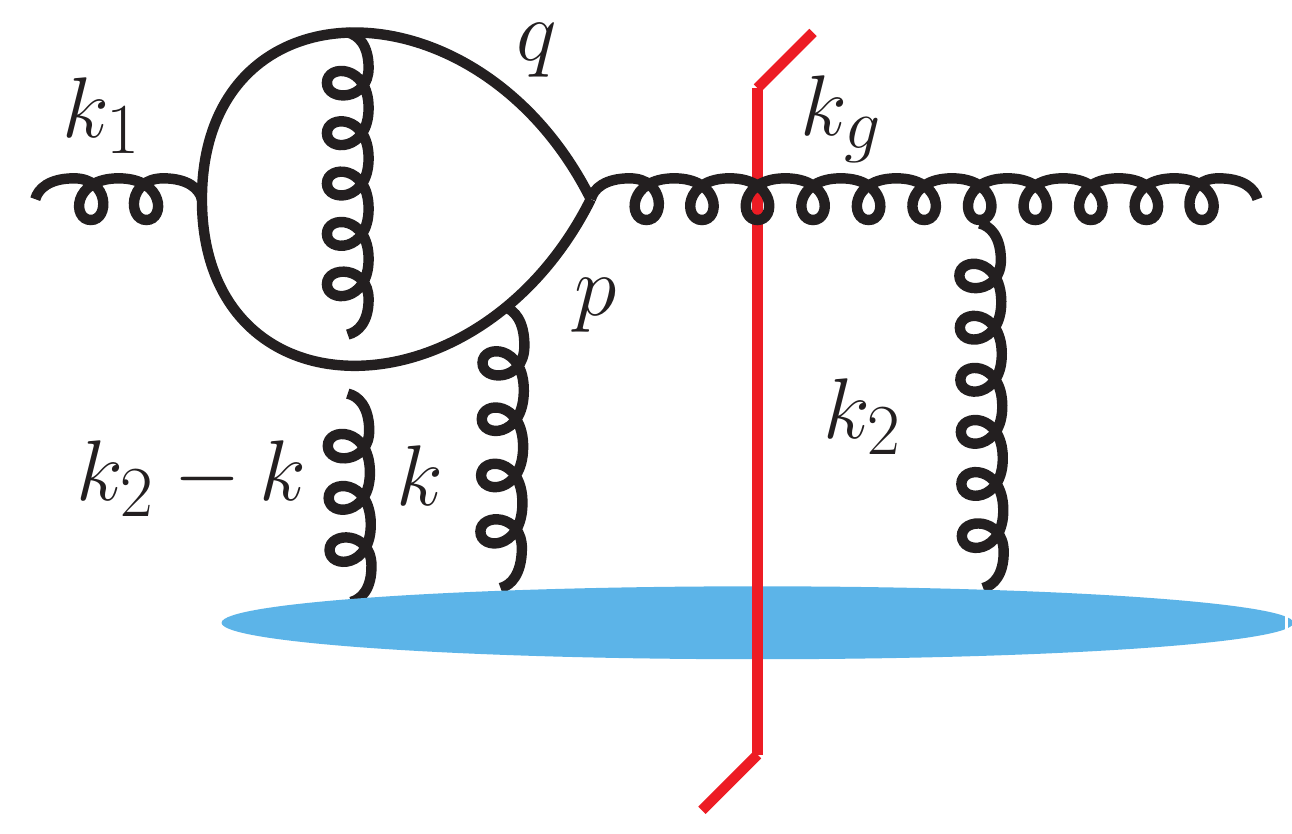}
  \end{center}
  \caption{An interference diagram that determines $S_{\alpha\beta}^{(0)}(k_1)$ in the virtual correction to the $g \to g$ channel. The vertical gluons denote Wilson lines arising from multiple scattering on the dense nucleus.}
  \label{fig:ggvirt}
\end{figure}
In the case of the $g \to gg$ channel there is a great simplification due to the fact that the purely gluonic contributions involve only adjoint Wilson lines, which are real, and therefore the odderon mechanism is absent. The only exception is the quark loop correction to the tree-level $g(k_1)\to g(k_g)$ amplitude, see Fig.~\ref{fig:ggvirt} which we compute below. Discarding immediately the dipole pieces, $S^{(0)}_{\alpha\beta}(k_1)$ takes the following form
\be
\begin{split}
S_{\alpha\beta}^{(0)}(k_1) & = \frac{1}{2P_p^+}(2\pi)\delta(k_1^+ - k_g^+) (-\rmi g^2) N_f N_c T_R \int_{q^+}\int_{\qp\kp\kp'}\int_{\xp\xp'\yp\yp'}\\
&\times\rme^{\rmi \kp\cdot\xp}\rme^{\rmi(\kAp - \kp)\cdot\yp}\rme^{-\rmi\kp \cdot\xp'}\rme^{-\rmi(\kAp - \kp')\cdot\yp'}\\
&\times\left[S_{qqg}(\xp,\yp,\xp') (-2k_1^+)d_{\alpha\mu}(k_g)\calT_{q\bar{q},\beta}^\mu(\kp) + S_{qqg}(\yp',\xp',\xp) (-2k_1^+)d_{\alpha\mu}(k_g)\calT_{q\bar{q},\beta}^{\mu\dag}(\kp')\right]\,.
\end{split}
\label{eq:S0gg}
\ee
where $q$ is the quark loop momentum and
\be
\calT_{q\bar{q}}^{\mu\beta}(\kp) \equiv \frac{1}{2q^+}\int_{q^-}{\rm tr}\left[\frac{\slashed{q}-\slashed{k}_g}{(q-k_g)^2 + \rmi \epsilon}\gamma^\mu \frac{\slashed{q}}{q^2 + \rmi \epsilon}\gamma^+ (\slashed{q}-\slashed{k}_g + \slashed{k}_1 + \slashed{k})\gamma^\beta \frac{\slashed{q}-\slashed{k}_g + \slashed{k}}{(q-k_g + k)^2}\gamma^+ \right]\,.
\ee
Similar to the computation in Sec.~\ref{sec:qqbar}, and according to \eqref{eq:wqqbar}, we are to evaluate the following combination
\be
\calH^{(\calG_{3T})}(\kp) - \left[\frac{\pd}{\pd k_1^\lambda}\calH^{(\tilde{g}),\lambda}(\kp,\khp)\right]_{k_1 = p_1}\,,
\ee
where we define
\be
\begin{split}
&\calH^{(\calG_{3T})}(\kp) \equiv \frac{1}{(2q^+)^2}\frac{1}{p_1^+} \epsilon^{n\alpha\beta S_\perp}\omega_{\alpha'\alpha}\omega_{\beta'\beta}(-2k_1^+)d_{\alpha'\mu}(k_g)\calT^\mu_{q\bar{q},\beta}(\kp)\,,\\
& \calH^{(\tilde{g}),\lambda}(\kp,\khp) \equiv \frac{1}{(2 q^+)^2} \left(g_\perp^{\beta\lambda} \epsilon^{\alpha \bar{n}n S_\perp} - g_\perp^{\alpha\lambda}\epsilon^{\beta \bar{n}n S_\perp}\right)(-2k_1^+)d_{\alpha\mu}(k_g)\calT^\mu_{q\bar{q},\beta}(\kp)\,.
\end{split}
\ee
A direct computation leads to
\be
\calH^{(\calG_{3T})}(\kp) - \left[\frac{\pd}{\pd k_1^\lambda}\calH^{(\tilde{g}),\lambda}(\kp,\khp)\right]_{k_1 = p_1} = \left[S_\perp^\lambda \frac{\pd}{\pd k_{1\perp}^\lambda}\left(4 (y^2 + \bar{y}^2)\frac{\vonp\times \vtwp}{\vonp^2 \vtwp^2}\right)\right]_{k_1 = p_1}\,,
\ee
where now $\vonp \equiv - \qp + \kgp - \bar{y}\khp - \kp$ and $\vtwp \equiv \qp - y \kgp$. But this is completely analogous to the result for the loop correction in Sec.~\ref{sec:virt} and so $S^{(0)}_{\alpha\beta}(k_1)$ vanishes after the $\qp$ integral.

\section{Conclusions and outline}
\label{sec:conc}

We have revisited the odderon mechanism for SSA in $p^\uparrow A$ originally suggested in \cite{Kovchegov:2012ga} at the quark level. At the hadron level this mechanism would involve the $g_T(x)$ distribution. We have considered the WW truncation of the full twist-3 polarized cross section and argued that in addition to $g_T(x)$ we also need to take into account the $g_{1T}^{(1)}(x)$ for a consistent computation. Our main finding is that under this truncation the polarized cross section vanishes exactly up to NLO for all possible partonic channels.

 %--- figure ---%
\begin{figure}
  \begin{center}
  \includegraphics[scale = 0.5]{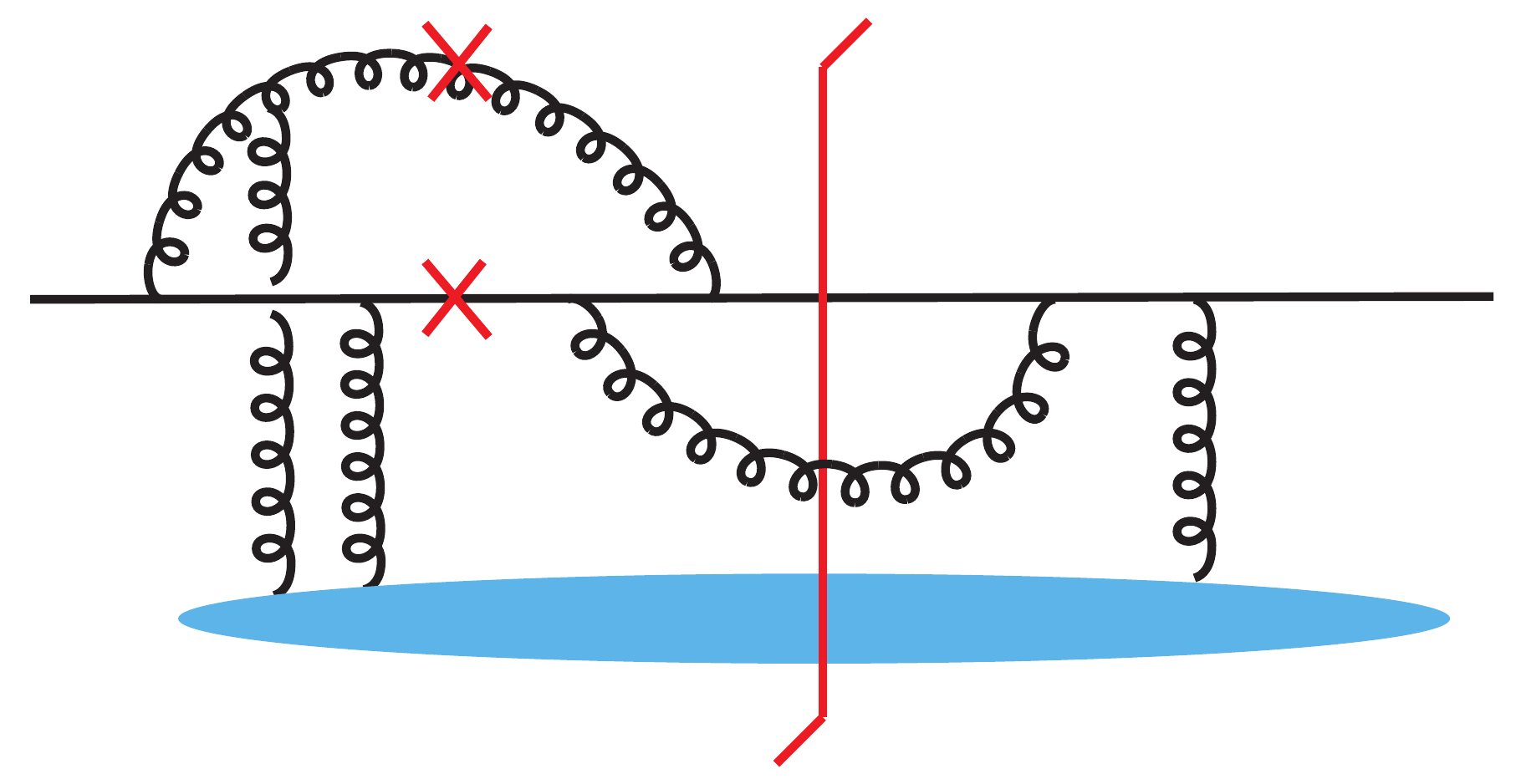}
  \end{center}
  \caption{A sample interference diagram appearing at NNLO. The crosses denote cut propagators that determine the imaginary part of the loop.}
  \label{fig:nnlo}
\end{figure}
It is natural to consider whether any of the above assumptions can be relaxed so that a non-zero contribution to SSA from the odderon mechanism may be found after all. One option is to go beyond the WW approximation, namely including the ETQS pieces in \eqref{eq:wmain}. Note the difference from the more conventional pole calculus -- here one needs to pick up the principal value of internal propagators so that the general functional forms of the ETQS functions would be required. Alternatively, one can consider the twist-3 FF mechanism, where we pick up the real part of the twist-3 FFs with the phase provided by the odderon. Once more, this in contrast to the conventional computations where the phase is supplied by the imaginary part of twist-3 FFs. Given that the current global fits constrain only the imaginary part of the twist-3 FFs \cite{Cammarota:2020qcw,Gamberg:2022kdb}, the phenomenological implications of this alternative would be worth exploring.

Another possibility would be to retain the WW approximation but compute the hard factor up to NNLO. While of course only an explicit computation can reveal whether the odderon appears at NNLO, we mention here a competing mechanism that is already known to appear at NNLO. The basic premise is very simple: at higher orders it is an imaginary part of the loop amplitude that can supply the phase. A specific NNLO contribution illustrating this is given in Fig.~\ref{fig:nnlo}, where the crosses denote cut propagators. Physically, the initial $q \to qg$ splitting occurs inside the target nucleus in the amplitude. The $qg$ system subsequently rescatters with a $t$-channel quark into the final state providing a phase with respect to the amplitude on the opposite side of the final state cut. Such final state rescattering is sometimes referred to as the lensing mechanism and was considered in \cite{Kovchegov:2020kxg}. In fact, this idea \cite{Brodsky:2002cx} is closely related to the very first estimate of SSA in perturbative QCD \cite{Kane:1978nd}. The computation in \cite{Kovchegov:2020kxg} was in the quark-diquark model. As a future work it would be important to consider this in the hybrid approach.

\acknowledgments

S.~B. thanks Yoshitaka Hatta for suggesting to work on the odderon mechanism for SSA. We thank Yoshitaka Hatta and Yuri Kovchegov for useful comments on the manuscript. S.~B., A.~K. and E.~A.~V. are supported by the Croatian Science Foundation (HRZZ) no. 5332 (UIP-2019-04).

\bibliographystyle{h-physrev}
\bibliography{references}

\end{document}